\documentclass[fleqn,usenatbib]{mnras}
\usepackage{newtxtext,newtxmath}

\usepackage[T1]{fontenc}

\DeclareRobustCommand{\VAN}[3]{#2}
\let\VANthebibliography\thebibliography
\def\thebibliography{\DeclareRobustCommand{\VAN}[3]{##3}\VANthebibliography}

\usepackage{graphicx}	
\usepackage{amsmath}	

\newcommand{\XY}[2]{\left[\textrm{#1/#2}\right]}
\newcommand{\FeH}{\XY{Fe}{H}}

\newcommand{\kms}{km\,s$^{-1}$}
\newcommand{\Teff}{T_\textrm{eff}}
\newcommand{\logg}{\log g}

\newcommand{\vmic}{v_\textrm{mic}}

\newcommand{\fspot}{f_\text{spot}}
\newcommand{\Bspot}{B_\text{spot}}
\newcommand{\Tspot}{T_\text{spot}}
\newcommand{\RHK}{R_\text{HK}'}
\newcommand{\geff}{g_\text{eff}}

\newcommand{\Felineone}{5247\,\AA}
\newcommand{\Felinetwo}{5374\,\AA}

\usepackage{scalerel,tikz}
\usetikzlibrary{svg.path}
\definecolor{orcidlogocol}{HTML}{A6CE39}
\tikzset{orcidlogo/.pic={
 \fill[orcidlogocol] svg{M256,128c0,70.7-57.3,128-128,128C57.3,256,0,198.7,0,128C0,57.3,57.3,0,128,0C198.7,0,256,57.3,256,128z};
 \fill[white] svg{M86.3,186.2H70.9V79.1h15.4v48.4V186.2z}
 svg{M108.9,79.1h41.6c39.6,0,57,28.3,57,53.6c0,27.5-21.5,53.6-56.8,53.6h-41.8V79.1z M124.3,172.4h24.5c34.9,0,42.9-26.5,42.9-39.7c0-21.5-13.7-39.7-43.7-39.7h-23.7V172.4z}
 svg{M88.7,56.8c0,5.5-4.5,10.1-10.1,10.1c-5.6,0-10.1-4.6-10.1-10.1c0-5.6,4.5-10.1,10.1-10.1C84.2,46.7,88.7,51.3,88.7,56.8z};
}}
\newcommand\orcidicon[1]{\href{https://orcid.org/#1}{\mbox{\scalerel*{
\begin{tikzpicture}[yscale=-1,transform shape]
\pic{orcidlogo};
\end{tikzpicture}
}{|}}}}

\title[Spectroscopic analyses of active stars]{A simple model for spectroscopic analyses of active stars}

\author[Nordlander, Baratella, Spina \& D'Orazi]{
T. Nordlander$^{\orcidicon{0000-0001-5344-8069}\,1,2}$\thanks{Stromlo Fellow}\thanks{E-mail: Thomas.Nordlander@anu.edu.au}, 
M. Baratella$^{\orcidicon{0000-0002-1027-5003}\,3}$, 
L. Spina$^{\orcidicon{0000-0002-9760-6249}\,4,5,2}$, 
and V. D'Orazi$^{\orcidicon{0000-0002-2662-3762}\,6,5}$ 
\\
$^{1}$Research School of Astronomy and Astrophysics, Australian National University, Canberra, ACT 2611\\
$^{2}$ARC Centre of Excellence for All Sky Astrophysics in 3 Dimensions (ASTRO 3D), Australia\\
$^{3}$ESO – European Southern Observatory, Alonso de Cordova 3107, Vitacura, Santiago, Chile\\
$^{4}$INAF, Astrophysical Observatory of Arcetri, Largo Enrico Fermi 5, 50125 Florence, Italy \\
$^{5}$INAF, Padua Astronomical Observatory, vicolo dell' Osservatorio 5, 35122 Padua, Italy\\
$^{6}$Department of Physics, University of Rome Tor Vergata, via della Ricerca Scientifica 1, 00133 Rome, Italy
}

\date{Accepted 2024 November 06. Received 2024 November 04; in original form 2024 May 11}

\pubyear{2024}

\begin{document}
\label{firstpage}
\pagerange{\pageref{firstpage}--\pageref{lastpage}}
\maketitle

\begin{abstract}
Spectroscopic analyses of young late-type stars suffer from systematic inaccuracies, typically under-estimating metallicities but over-estimating abundances of certain elements including oxygen and barium. 
Effects are stronger in younger and cooler stars, and recent evidence specifically indicates a connection to the level of chromospheric activity.
We present here a two-component spectroscopic model representing a non-magnetic baseline plus a magnetic spot, and analyse the resulting synthetic spectra of young solar analogues using a standard spectroscopic technique. 
For a moderately active star with solar parameters and chromospheric activity index $\log \RHK = -4.3$ ($\sim 100$\,Myr), we predict that $\FeH$ is underestimated by 0.06\,dex while $\vmic$ is overestimated by 0.2\,\kms; for higher activity levels we predict effects as large as 0.2\,dex and 0.7\,\kms. Predictions are in agreement with literature data on solar twins, and indicate that the model is a plausible explanation to the observed effects. The model is simple enough that it can be included in spectroscopic packages with only changes to the underlying spectrum synthesis modules, if a $\log \RHK$ value is provided. 
\end{abstract}

\begin{keywords}
stars: activity -- stars: atmospheres -- stars: magnetic fields -- sunspots
\end{keywords}

\section{Introduction}
Young solar-type stars are characterised through rapid rotation, relatively strong magnetic fields and associated high levels of variability, together called `activity'. 
These phenomena are connected, as rotation drives the dynamo that generates magnetic fields, and together they deposit energy in the chromosphere and corona \citep{wedemeyer-bohm_magnetic_2012} which emit in the cores of strong lines \citep[e.g.,][]{leighton_observations_1959} as well as in (soft) X-rays \citep{zuckerman_young_2004}; the magnetic stellar wind in turn produces a torque that dissipates angular momentum \citep{matt_mass-dependence_2015}.

The past two decades have amassed evidence of a correlation between activity levels and distortions in spectroscopic analyses. 
\citet{morel_photospheric_2003,morel_photospheric_2004} found correlations between chromospheric activity and photometric colours as well as an apparent deficit in iron abundances and a corresponding enhancement in abundance ratios for several elements in their most active (coolest) stars. 
A particularly strong signal is the so-called barium puzzle, where \citet{dorazi_enhanced_2009} found a trend of increasing Ba abundance with decreasing age, where dwarf stars in very young clusters (30--50\,Myr) and active giants in clusters of 700\,Myr were found to be enhanced by as much as 0.6\,dex compared to solar. 
Following that study, other works validated the observed trend in Ba abundances as a function of age for both clusters and field stars \citep{yong2012, dasilva2012, mishenina2015, reddy2015, reddy2017, magrini2018}. There is also evidence of a connection not only with age but also the level of stellar activity \citep{baratella_gaia-eso_2021}.
Further evidence that young stars may be exhibiting distorted abundances comes from studies of star forming regions and young open clusters, which surprisingly exhibit subsolar metallicities for a range of locations in the Galaxy \citep[e.g.,][]{cunha_chemical_1998,santos_chemical_2008,spina_gaia-eso_2017}. 
Similarly, \citet{kos_galah_2021} found complex abundance variations in the Orion complex ($<20$\,Myr) that correlate well with effective temperature ($\Teff$), with very small star-to-star variations regardless of spatial location in the complex.  

Measurements from particular ionic stages of an atom may be distorted by factors related to youth or activity. For example, dwarfs in young open clusters exhibit a significant deviation from ionisation equilibrium where lines of the ionised species appear correlated with $\Teff$. In the $\sim150$\,Myr Pleiades cluster, the coolest stars have been found to experience overestimated abundances of ionised iron by nearly 1\,dex \citep[][]{schuler_fe_2010}. 
Another finding is that spectroscopic estimates of the microturbulent velocity ($\vmic$) are often anomalously high for young stars. For example, \citet{james_fundamental_2006} reported measurements in star forming regions ($<10$\,Myr) as high as 2.5\,\kms, along with significantly sub-solar metallicities in the coolest stars. 
As a potential solution to the issue of anomalously high $\vmic$ values, \citet{baratella_gaia-eso_2020} presented a methodology based on measurements from lines of titanium rather than iron, leading to significantly lower $\vmic$ values as well as abundances more in line with expectations for young solar-neighbourhood clusters. They also found evidence that the discrepancy between $\vmic$ derived from lines of titanium, as compared to iron, depended not only on age but also with the level of stellar activity.

A separate line of evidence has come from spectroscopic time series, where variations in spectroscopic quantities have found to correlate with variations in chromospheric activity. 
\citet{flores_discovery_2016} studied how the chromospheric activity indicator $S$ correlated with other spectral features in the solar analog HD 45184. While they found a straightforward correspondence with emission in the Balmer line cores, they also discovered an amplification for a number of strongly saturated singly ionised lines of iron, titanium and barium, as well as the inner wings of the Balmer lines. 
For the young solar twin HIP 36515, \citet{yana_galarza_effect_2019} found small but significant evidence of variations in spectroscopic parameters as well as in the equivalent widths of a iron lines. Foremost, they found that the variations were most pronounced for lines that formed high in the photosphere \citep[in line with results from][]{reddy2017}, and for those with large magnetic sensitivity. 
Using a standard list of iron lines, their spectroscopic analysis resulted in small but significant correlations between the chromospheric activity level and the inferred values of $\Teff$ and the metallicity $\FeH$ (which decreased) and $\vmic$ (which increased). 
\citet{spina_how_2020} performed an extended analysis on a large sample of 211 solar analogue stars covering chromospheric activity indices $\log \RHK = -5.4$ to $-4.3$, that exhibited variations in $\log \RHK$. For individual stars in their sample with activity indices $\log \RHK > -5.0$, stellar parameters and line equivalent widths were found to vary significantly, in line with results from \citet{yana_galarza_effect_2019}. Variations correlated both with the value of $\log \RHK$ and with the variation in $\log \RHK$.  

The overall result is that increasing stellar activity leads to an overestimate of $\vmic$ and therefore an underestimate of $\FeH$, in spectroscopic studies based on iron lines. This indicates that stellar activity is a good candidate to explain the various issues found in young stars. While it appears possible to derive stellar parameters from a selection of lines that are largely insensitive to stellar activity \citep{baratella_gaia-eso_2020}, this doesn't solve the problem of measuring abundances from species where no activity-insensitive lines exist. Thus, a spectroscopic model that takes into account the effects of magnetic cool spots appears necessary. 

In this paper, we present a spectroscopic model for the effects of magnetic cool spots (Sect.~\ref{sec:methods}) and estimate the biases this introduces on traditional spectroscopic analyses of young solar analogues. 
Results of a parameter study for solar-type stars are presented in Sect.~\ref{sec:results}. 
In Sect.~\ref{sec:discussion} we discuss empirical estimates of the underlying parameters, and connect these to a single observable -- the activity index $\log \RHK$. We present predicted effects on spectroscopic parameters for solar-type stars, and conclude in Sect.~\ref{sec:conclusions} that these are in agreement with observations.

\section{Methods} \label{sec:methods}
We mimic the spectroscopic effects of stellar activity using a two-component photospheric model. The model represents a non-magnetic reference photosphere plus a magnetic spot that causes Zeeman splitting. The spot has identical spectroscopic parameters to the reference component, aside from a different effective temperature $\Teff$ and magnetic field $B$. 
We analyse these synthetic spectra as if they represented real data, to estimate effects on stellar parameters that would be seen in analyses of active stars.

\subsection{Spectrum synthesis code}
We have extended the spectrum synthesis code TurboSpectrum \citep[v15.1;][]{alvarez_near-infrared_1998,plez_turbospectrum:_2012} to take into account Zeeman splitting of spectral lines, following \citet{stenflo_solar_1994}. 
For Zeeman triplets, magnetic fields of strength $B$ produce a splitting of
\begin{equation}
    \Delta v_\text{Z} = \frac{\mu_\text B}{h} \geff\, \lambda B \approx 0.700\,\text{\kms} \, \geff \, (\lambda \,/\,5000\,\text{\AA}) \, (B \,/\, \text{kG}),
    \label{eqn:magsplit}
\end{equation}
where $\mu_\text B \approx 5.79 \times 10^{-9}\,\text{eV}\,\text{G}^{-1}$ is the Bohr magneton and $\geff$ is the effective Land\'e factor.
In the optical, magnetically sensitive lines (with $\geff \gtrsim 1$) thus experience Zeeman splitting of the order a few \kms\ within regions of strong magnetic fields such as star spots. Such splitting may be significantly broader than $\vmic$, and can in a similar way act to de-saturate spectral lines.

Calculations are an extension of those presented in \citet{nordlander_lowest_2019}, and we likewise produce synthetic spectra using a grid of MARCS model atmospheres \citep{gustafsson_grid_2008}. We use atomic data from the Vienna Atomic Line Database \citep[VALD3;][]{piskunov_vald:_1995, kupka_vald-2:_1999, ryabchikova_major_2015}, which provides Land\'e factors for tens of thousands of atomic lines in the optical region. We explicitly calculate anomalous Zeeman splitting for lines whenever Land\'e factors are available for both states of a transition.

Additionally, we take into account molecular lines as background opacity but do not apply Zeeman splitting here as our molecular databases lack the necessary Land\'e factors. Although these are possible to retrieve \citep{berdyugina_successful_2000,afram_molecules_2015}, calculations are involved and beyond the scope of this work as we focus on the traditional analysis of atomic lines in solar-type stars with molecular lines primarily contributing background opacity.
We include a total of 15 million molecular lines, representing the molecules AlH, C$_2$, CaH, CH, CN, CO, CrH, FeH, H$_2$O, MgH, NH, OH, SiH, SiO, SiS, TiO, VO and ZrO. A complete list of references is provided in Table~\ref{tbl:molecules}. In this work, optical atomic lines that are unblended in the solar spectrum have been selected and thus only molecules that appear in cool spots are of importance, meaning primarily CN and TiO. In the case of TiO, we note that magnetic splitting is only ever significant for very low ($J < 10$) rotational levels \citep{berdyugina_molecular_2002}.

\subsection{Verification on sunspots}
We find that our synthetic spectra are able to reasonably well reproduce the spectrum of a sunspot umbra \citep{wallace_atlas_1999}. We note that while this sunspot has an excellent spectrum of high resolution and high signal to noise, it is not further identified or analysed by the authors and its reference properties such as the location on the solar surface are not known to us. 
As an approximation of the line formation in a single sunspot, we use a flux (full stellar disc) spectrum with solar parameters except for the choice of a lower effective temperature. We thus neglect the potential influence of magnetic pressure on the atmospheric structure as well as potential deviations from non-magnetic hydrostatic equilibrium and viewing angle effects \citep[see, e.g.,][]{rempel_numerical_2012}. 

\begin{figure}
    \centering
    \includegraphics[width=\columnwidth]{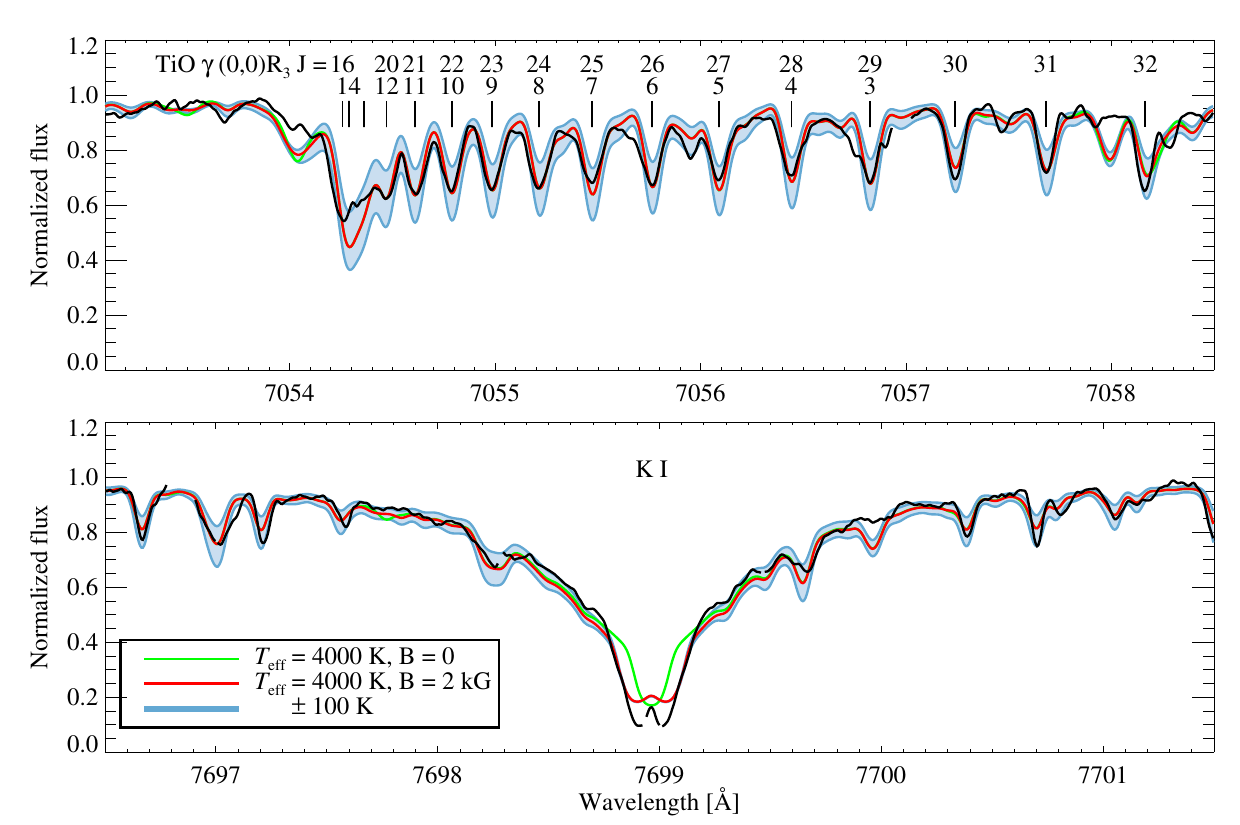}
    \caption{A high-resolution umbral sunspot atlas (black solid line) compared to synthetic spectra with and without magnetic fields. 
    The TiO molecular $\gamma$-system $v=0$--0 R$_3$ band head (top) exhibits a strong temperature sensitivity, but no magnetic splitting. Synthetic spectra are shown with (red line) and without (green line) magnetic fields, and with a variation in $\Teff$ (shaded blue area). Molecular TiO lines are identified by their rotational quantum numbers which run from right to left for $J<16$ and left to right for $J>16$. 
    The \ion{K}{i} resonance line (bottom) exhibits strong Zeeman doublet splitting which allows determining the magnetic field in the umbra of this sunspot.}
    \label{fig:verification}
\end{figure}

As illustrated in Fig.~\ref{fig:verification}, we find a good match to the temperature sensitive TiO $\gamma$ bandhead at 7054\,\AA\ with $\Teff = 4000$\,K, which is not unreasonable for a typical sunspot umbra \citep[e.g.,][]{sobotka_high-resolution_1993}. 
Besides the shape of the bandhead itself, the shape and depth of these molecular lines is well reproduced by our synthetic spectrum. 
No magnetic splitting is evident except for possibly the lines with lowest $J$, which are expected to exhibit the largest Land\'e factors \citep[$g > 1$][]{berdyugina_molecular_2002}. 
We also find an excellent match to the wings of the \ion{K}{i} 7699\,\AA\, resonance line at the same temperature. The core of this line is not temperature sensitive but exhibits significant Zeeman doublet splitting. The observed separation matches well our prediction with $B \sim 2$\,kG, which again is a reasonable value for sunspot umbrae.

\subsection{Synthetic spectrum grid}
We generate synthetic spectra representing stars with solar stellar parameters and varying levels of magnetic activity. 
These are represented by a two-component model with a non-magnetic `quiet' component plus a single magnetic `spot' component. Similar models have been used in the literature \citep[e.g.,][]{basri_physical_1988,rueedi_magnetic_1997,kochukhov_hidden_2020}.
We define a spot filling factor $f$ and calculate the composite spectrum
\begin{equation}
    F_\text{composite} = \fspot F_\text{spot} + (1-\fspot) F_\text{quiet}.
\end{equation}
The synthetic spectra differ in $\Teff$ and in magnetic field strength $B$, but otherwise use the same stellar parameters: surface gravity $\logg=4.44$, metallicity $\FeH=0.0$ and microturbulence $\vmic=1.0$\,\kms. 
For the quiet component, we set an effective temperature $T_\text{quiet} = 5771$\,K and $B_\text{quiet} = 0$. 

We note that this approach leads to a significant decrease in luminosity compared to a not-spotted star of the same $\Teff$, by upwards of 30\,\%, but argue below that various approaches around this used in the literature would have a rather small impact on our model. 
For example, the luminosity can be explicitly conserved by varying $T_\text{quiet}$ while conserving $\Teff^4 = \fspot \Tspot^4 + (1-\fspot) T_\text{quiet}^4$ \citep[e.g.,][]{somers_older_2015}. This approach was used, e.g., by \citet{wilson_stellar_2023}, who found that even with no magnetic fields, the presence of spots and its star-to-star variation may introduce a scatter in the inferred metallicities at the 0.05\,dex level.
Studies of both the Sun and solar analogues also indicate that the luminosity is not necessarily conserved on short timescales, as periods with large sunspots or of increased chromospheric activity correspond to significant reduction in brightness \citep[e.g.][]{willson_observations_1981,hall_activity_2009}. 

It's also possible that magnetic pressure either offsets or adds to the gas pressure. Magneto-hydrodynamic studies have found a depression in gas pressure, compared to the non-magnetic case, by some 30\,\%; this lower gas density decreases the opacity such that the optical surface forms deeper, where both pressure and temperature are relatively higher \citep[see][for details]{rempel_numerical_2012}. We note that our verification on the umbra of a sunspot using pressure-sensitive features did not obviously require any such changes (see Fig.~\ref{fig:verification}).

For the spot components, we consider synthetic spectra over the ranges $\Teff = 4771$--5771\,K, $B = 0$--3\,kG. We generate composite spectra for a wide range of spot filling factors, $\fspot = 0$--0.5. 
An extensive grid of synthetic spectra covering FGK-type stars with spots is generated for use in future work.

\subsection{Spectroscopic measurements on synthetic spectra}
We perform spectroscopic measurements on our synthetic spectra representing spotted stars, treating these as if they were real stars. We analyse spectra using a strictly differential line-by-line analysis method, as outlined by \citet[][and references therein]{spina_how_2020}.

Briefly, we created a line list taking the one from \citet{melendez_18_2014} as reference.  It comprises a set of 94 lines of \ion{Fe}{i}, 18 of \ion{Fe}{ii}, as well as 59 lines of other metals including Sc, Ti, V, Cu, Y and Ba (in light of the results reported in \citealt{baratella_gaia-eso_2021}). The lines have been selected based upon their suitability for analysis in solar type stars, however with no consideration of their magnetic sensitivity. In practice, their effective Land\'e factors range from $\geff = 0$ to 3. 

The equivalent widths (EWs) of each line were measured with the \texttt{smhr}\footnote{\url{https://github.com/andycasey/smhr}} and \texttt{stellar diff}\footnote{\url{https://github.com/andycasey/stellardiff}} codes, which allow the user to define masks for continuum setting around the spectral line of interest and to measure the EWs, respectively. The masks are created considering regions of continuum completely free from strong absorption lines, within a spectral window varying from 3 to 5 \AA\ around the line of interest. Once the masks are created, \texttt{stellar diff}  applies the same continuum mask to all the spectra, fitting the lines with a Gaussian profile. 

After measuring EWs, we derive stellar parameters for each composite synthetic spectrum using the \texttt{qoyllur-quipu}\footnote{\url{https://github.com/astroChasqui/q2}} code (q$^2$, \citealt{2014ramirez}). This publicly available tool is a python wrapper of MOOG (version 2019) that automatically derives the stellar parameters ($\Teff, \logg$, $\vmic$, and $\FeH$) through a line-by-line differential analysis. In particular, the parameters and abundances are computed with respect to a reference star, which in our case is the synthetic quiet solar spectrum, to minimise the impact of model uncertainties and errors in the atomic data. We refer the reader to \cite{melendez_18_2014} and \cite{2014ramirez} for a more detailed description of the technique. 

This results in an estimate of the stellar parameters and how these would vary in real stars as a function of our model parameters representing stellar activity: $\fspot$, $\Tspot$ and $\Bspot$.

\subsection{Photometric predictions}
We calculate synthetic photometry from our model spectra of spotted stars. 
Calculations used transmission curves for the Johnson-Cousins system from \citet{bessell_standard_2005} and for 2MASS from \citep{cohen_spectral_2003}, to calculate the $B$, $V$ and $K_s$ magnitudes. Combinations of these are commonly used in the literature to estimate stellar temperatures. 
For the colour-temperature transformation, we used the commonly used empirical calibrations from \citet{casagrande_absolutely_2010}, and shifted our predicted colours so that the non-spotted reference model reproduces the solar nominal $\Teff$ value, 5771\,K.

\section{Results} \label{sec:results}
We investigate variations in inferred stellar parameters for young solar analogues with spot coverage fraction ($\fspot$), spot magnetic field strength ($\Bspot$) and spot temperature ($\Tspot$). Specifically, we investigate variations in the inferred effective temperature ($\Teff$), surface gravity ($\logg$), metallicity ($\FeH$) and microturbulence ($\vmic$), as well as the spectral lines \ion{Fe}i \Felineone\ and \Felinetwo. 
The two reference spectral lines are selected to have similar line strength (about 60\,m\AA) and with similarly strong magnetic sensitivity, but with very different temperature sensitivity. \ion{Fe}i \Felineone\ has $\geff = 2.0$, $E_\text{low} = 0.09$\,eV, while \ion{Fe}i \Felinetwo\ has $\geff = 1.67$, $E_\text{low} = 4.47$\,eV.
As we have varied the three stellar activity markers independently, we illustrate them here one at a time.

\begin{figure*}
    \centering
    \includegraphics[width=\textwidth]{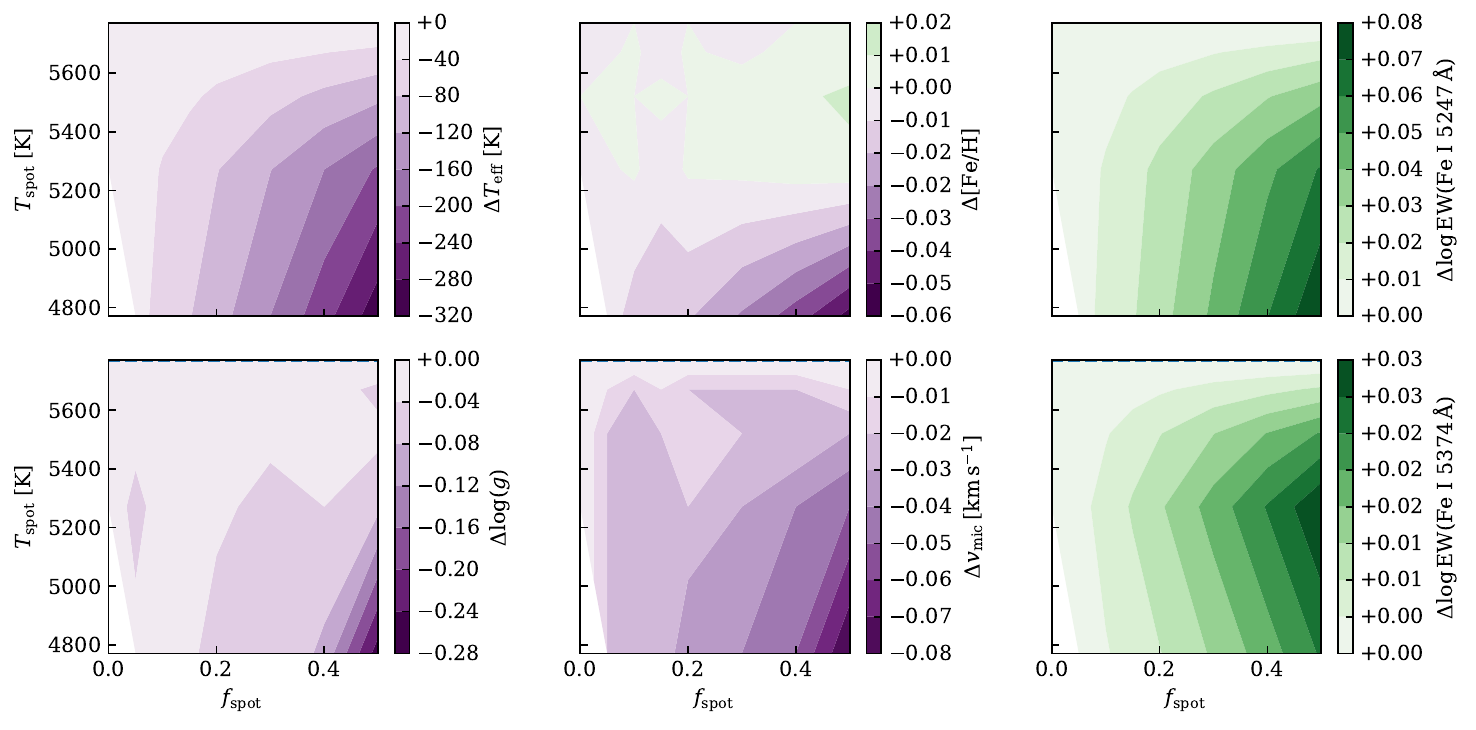}
    \caption{Predicted changes to stellar parameter measurements in young solar analogues with variations in spot coverage fraction ($\fspot$) and spot temperature ($\Tspot$) when neglecting magnetic fields. 
    The inferred stellar parameters are the effective temperature ($\Teff$), surface gravity ($\logg$), metallicity ($\FeH$) and microturbulence ($\vmic$). The rightmost panels indicate logarithmic changes to the strengths of two \ion{Fe}i lines with low (top) and high (bottom) excitation energy but similar line strength and magnetic sensitivity (see text for details). The somewhat blotchy behaviour for $\logg$, $\FeH$ and $\vmic$ is due to imperfections in convergence of our spectroscopic fitting method; seemingly random variations at these smallest scales should not be interpreted as real. }
    \label{fig:frac-T-nonmagnetic}
\end{figure*}

Figure~\ref{fig:frac-T-nonmagnetic} illustrates effects on stellar parameters when neglecting magnetic fields but varying the spot coverage ($\fspot$) and temperature ($\Tspot$). 
We find a straightforward correspondence between $\Tspot$ and the result for $\Teff$ (effects as large as 300\,K), and likewise a significant strengthening of \ion{Fe}{I} \Felineone\ with decreasing $\Tspot$ (strengthening by as much as 0.08\,dex or 13\,m\AA). \ion{Fe}I \Felinetwo\ is only weakly sensitive to temperature and does not vary significantly with $\Tspot$. 
Other stellar parameters do not exhibit straightforward variations with $\Tspot$. This indicates that our spectroscopic analysis is rather immune to these changes. In other words, there is no significant cross-talk between the surface temperature and these other parameters except in the most extreme case (at the largest $\fspot$, smallest $\Tspot$ value), but rather effects are captured by the estimated $\Teff$ value.

\begin{figure*}
    \centering
    \includegraphics[width=\textwidth]{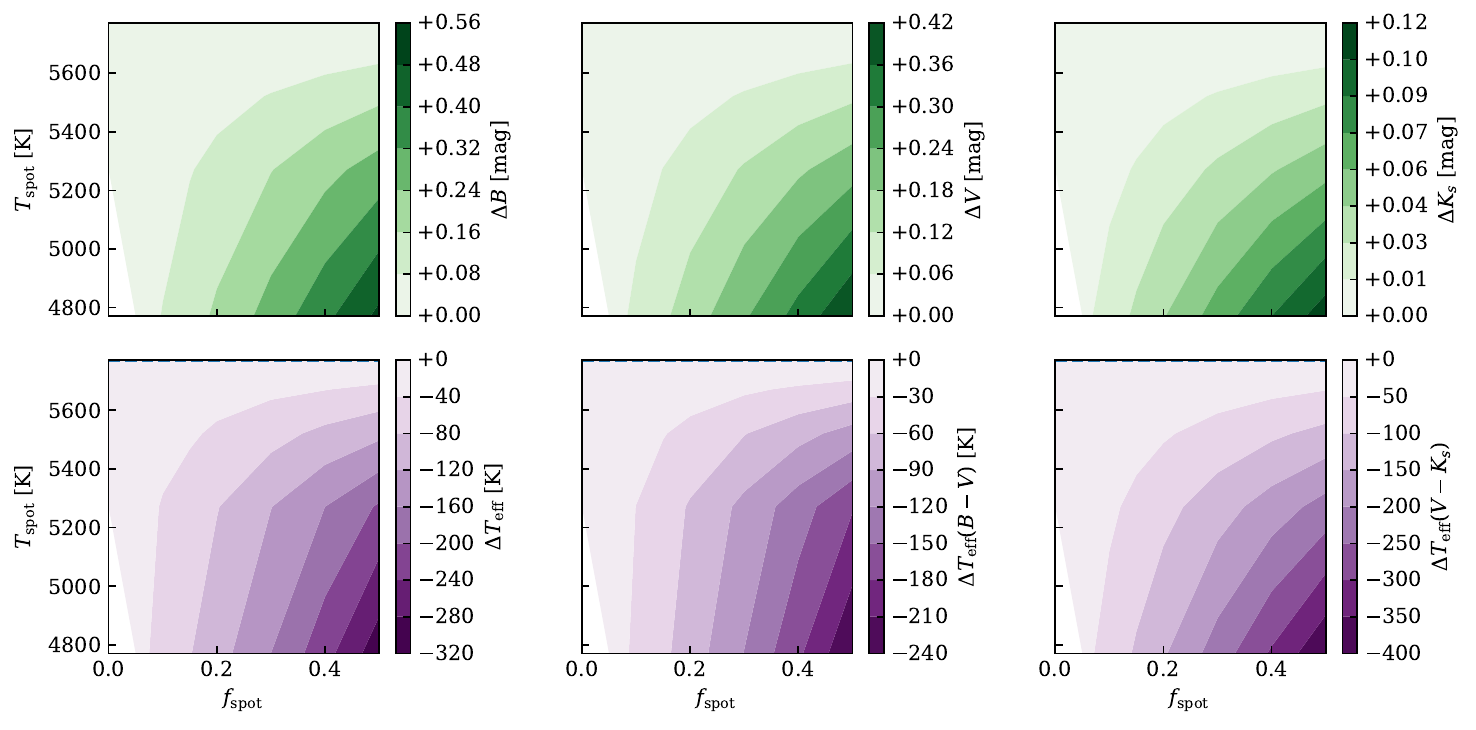}
    \caption{Predicted changes to the photometric magnitudes and resulting temperature estimates in young solar analogues with variations in spot coverage fraction ($\fspot$) and spot temperature ($\Tspot$) when neglecting magnetic fields. Shown are the direct effects on the $B$, $V$ and $K_s$ magnitudes, the spectroscopic effective temperature ($\Teff$; same as in Fig.~\ref{fig:frac-T-nonmagnetic}), and photometric $\Teff$ estimates from $B-V$ and $V-K_s$.}
    \label{fig:temperatures}
\end{figure*}

We also estimate effects on the photometric magnitudes when neglecting magnetic fields but varying the spot coverage ($\fspot$) and temperature ($\Tspot$), shown in Fig.~\ref{fig:temperatures}. 
Effects are straightforward, with effects as large as $0.56$\,mag in $B$, $0.42$\,mag in V, and 0.12\,mag in $K_s$. 
The corresponding effects to $\Teff$ as estimated from the $B-V$ and $V-K_s$ colours show a similar behaviour to the spectroscopic $\Teff$, resulting primarily from the change of the spectral energy distribution and with only a minor secondary effect from the varying line blanketing. 

\begin{figure*}
    \centering
    \includegraphics[width=\textwidth]{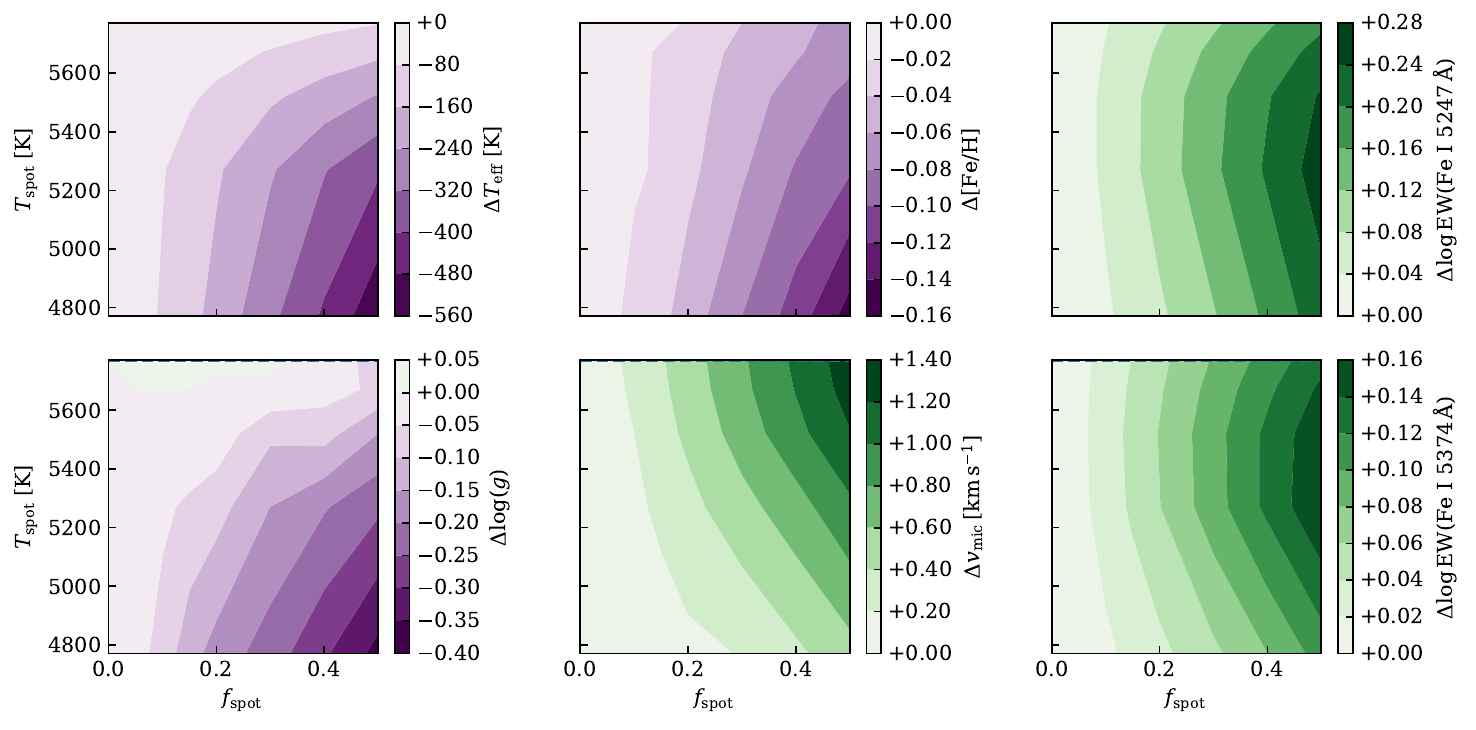}
    \caption{Predicted changes to stellar parameter measurements with variations in spot coverage fraction ($\fspot$) and spot temperature ($\Tspot$) when assuming a spot magnetic field of $\Bspot = 3$\,kG. Labels and parameters are the same as in Fig.~\ref{fig:frac-T-nonmagnetic}. 
    }
    \label{fig:frac-T}
\end{figure*}

In Fig.~\ref{fig:frac-T}, we investigate the effects on stellar parameters when spots are strongly magnetic, with $\Bspot = 3\,\text{kG}$, but leaving the `quiet' stellar surface nonmagnetic. 
With this setup, we find a straightforward variation in some stellar parameters as $\fspot$ increases. 
We find that the strong magnetic field amplifies effects on $\Teff$, and furthermore results in a reduction of $\FeH$ and an increase in $\vmic$, as well as an increase in the strengths of both \ion{Fe}i lines.
When varying $\Tspot$, we find a stronger impact on $\Teff$ than in the non-magnetic case. 
We find a weakened correspondence between $\Tspot$ and the strength of the \ion{Fe}i lines, with a non-monotonic variation. Rather, the lines strengthen significantly due to magnetic amplification, by up to 0.3\,dex (60\,m\AA) for \ion{Fe}i \Felineone\ and 0.16\,dex (30\,m\AA) for \ion{Fe}i \Felinetwo. 
Effects on $\FeH$ and $\logg$ essentially scale linearly with $\Tspot$ and $\fspot$, and grow as large as $-0.16$\,dex and $-0.4$\,dex. The behaviour of $\vmic$ is more complex, where the largest effects ($+1.4$\,\kms) are found for spots that are magnetic but with no temperature variation, while cool spots result in effects as large as $+0.7$\,\kms. 
Effects on the photometric $\Teff$ are not shown as they are effectively identical to the non-magnetic case. 

We note that the spectral lines selected for this comparison are strongly magnetically sensitive and therefore prone to `magnetic intensification' \citep{babcock_magnetic_1949}. As indicated by the lack of correspondence between effects on $\FeH$ and the strength of these individual lines, there isn't necessarily a direct correspondence between the altered strength of the spectral lines and their inferred abundance due to the variation of other stellar parameters.

Ample data are available on the coverage and properties of sunspots. Typical temperatures are lower than the surrounding solar surface, with a difference between spot temperatures and the effective temperature of $\Delta T \equiv \Tspot - \Teff \sim -1000$ to $-2000$\,K for the umbra of spots and $\Delta T \sim -300$\,K for the penumbrae \citep{solanki_sunspots_2003}. 
Suites of 3D magneto-radiation-hydrodynamic (MRHD) simulations of the solar photosphere with strong magnetic fields produce spots with typical temperatures approximately 1000\,K cooler than the photosphere \citep{heinemann_mhd_2007,rempel_penumbral_2009,rempel_radiative_2009,rempel_penumbral_2011,rempel_numerical_2012}. 
For young solar analogues, inversion methods have indicated large spots that exhibit similar temperature contrasts \citep[see the review by][]{berdyugina_starspots_2005}. 
For example, the $\sim50$\,Myr solar twin EK Dra exhibits a solar-like activity cycle and large-scale spots with umbral temperature contrasts $\Delta T \sim -1000$ to $-2000$\,K  \citep{rosen_magnetic_2016,waite_magnetic_2017,jarvinen_mapping_2018}. 
Similarly, temperature contrasts of $\Delta T \approx -2000$\,K were found for a number of $\sim 20$--30\,Myr solar analogues, including the IC 2602 member HD 307938 \citep{marsden_doppler_2005}, HD 29615 \citep{waite_magnetic_2015} and HIP 89829 \citep{perugini_evolution_2021}. 
Another $\sim30$\,Myr solar analog star, V1358 Ori, exhibits both cool spots with $\Delta T \sim -1500$\,K and hot spots with $\Delta T \sim +350$\,K \citep{kriskovics_magnetic_2019}. 
It therefore appears reasonable to assume that large cool spots are predominantly featured on the surfaces on young solar-type stars, and that a temperature contrast for the cool spots of $\Delta T = - 1000$\,K is not unreasonable. 
In the rest of this section and the following, we focus on this scenario.

\begin{figure*}
    \centering
    \includegraphics[width=\textwidth]{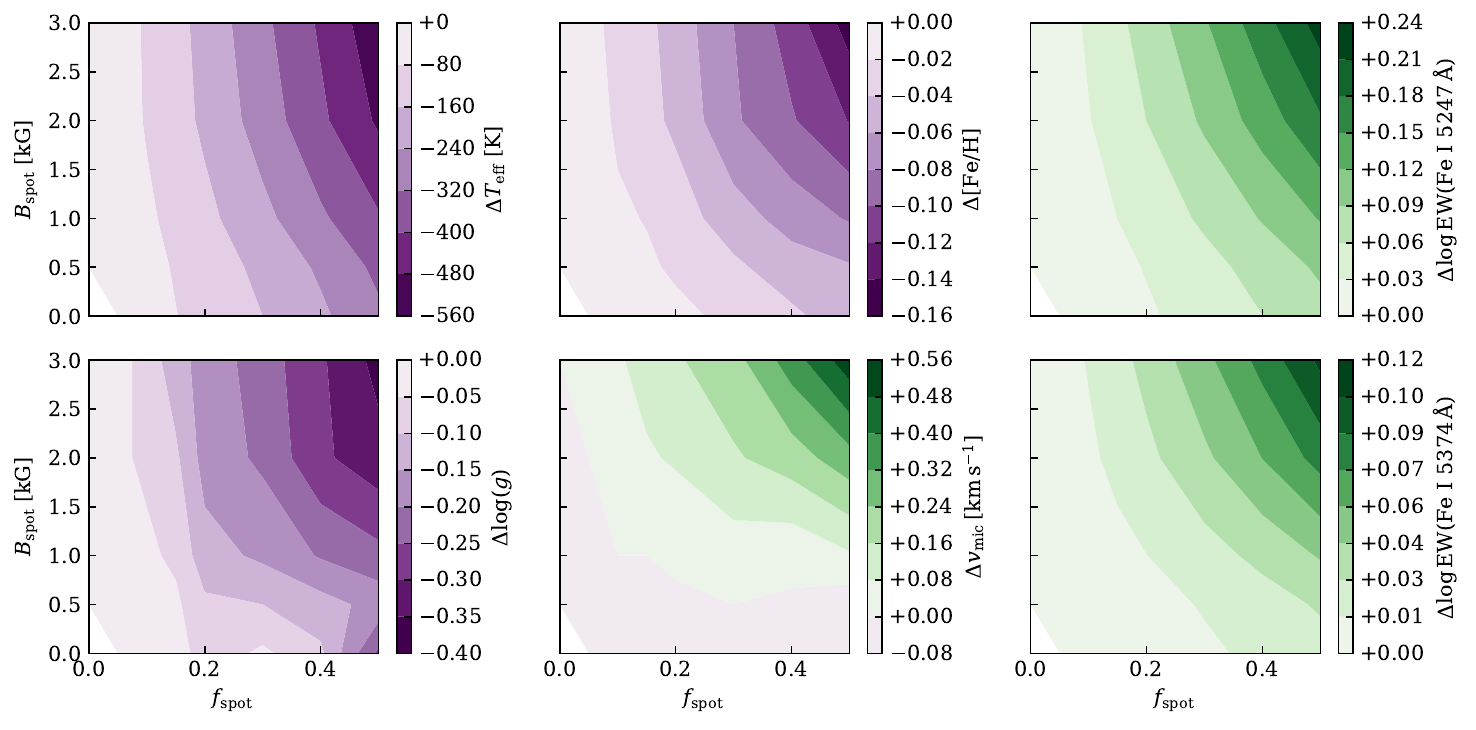}
    \caption{Predicted changes to stellar parameter measurements with variations in spot coverage fraction ($\fspot$) and spot magnetic field strength ($\Bspot$). Calculations assume a spot temperature $\Tspot = 4771$\,K, i.e., 1000\,K below $\Teff$. 
    Effects on stellar parameter measurements grow stronger with increasing $\fspot$ and $\Bspot$, with generally negative sign for $\Teff$, $\logg$ and $\FeH$, but positive sign for $\vmic$ and the strength of the \ion{Fe}i lines. }
    \label{fig:frac-B}
\end{figure*}

We select $\Tspot = 4771$\,K (i.e., $\Tspot = \Teff - 1000$\,K) for our test case illustrated in Fig.~\ref{fig:frac-B}. 
With this choice, we find a direct relation between $\Bspot$ and stronger effects on the inferred parameters. 
Clearly, the effect of magnetic intensification is directly responsible for the bulk of effects on all stellar parameters.  
We note that the \ion{Fe}i lines were selected because they are saturated and magnetically sensitive, with similar Land\'e $\geff$ values, and therefore both exhibit strong magnetic amplification. 
We note also that despite this amplification, we see a decrease in $\FeH$; this is largely due to the decrease in $\Teff$.
As previously indicated, varying the magnetic field strength has effectively no impact on the estimated photometric magnitudes and resulting $\Teff$ estimates, and these plots are therefore not shown.

\section{Discussion} \label{sec:discussion}
We have shown in Sect.~\ref{sec:results} that our two-zone stellar model comprising a non-magnetic `quiet' component combined with a cool magnetic component together yield significant variations in the inferred spectroscopic stellar parameters for young solar analogues. In particular, we find that models with large coverage of cool spots and with a strong magnetic field yield significant variations in all stellar parameters of a magnitude that is possible to measure in high-precision spectroscopic work. 
In order to interpret spectroscopic data in the literature, it would be useful to predict the spectroscopic effects purely from external indicators. Predictions from this model could be incorporated into spectroscopic analyses with no additional free parameters, beyond those inherent to the model itself.

\subsection{Inferring spot properties in active stars} \label{sec:translation}
Spots cannot normally be resolved on the surface of stars through direct imaging. We therefore need a means of translating our theoretical models (parametrised by $\fspot$, $\Bspot$ and $\Tspot$) into observables that are commonly available for solar-type stars.
The chromospheric activity index $\RHK$ is such a suitable observable, but unfortunately lacks a direct connection to our theoretical parameters. 

\begin{figure}
    \centering
    \includegraphics[width=\columnwidth]{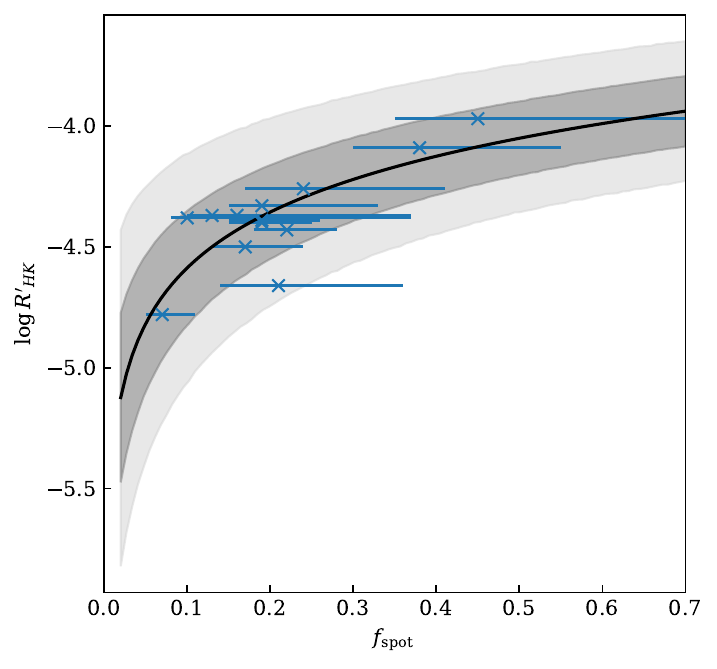}
    \caption{Empirical relationship between spot coverage fraction ($\fspot$) and activity index ($\log \RHK$) for solar analogue stars from \citet{kochukhov_hidden_2020}. 
    A power law fit is overplotted, constrained for activity indices in the range $-4.8 < \log \RHK < 4.0$. The $1\sigma$ and $2\sigma$ confidence regions are indicated by the vertical dark and light grey shaded regions. 
    } 
    \label{fig:frac-lRHK}
\end{figure}

\citet{kochukhov_hidden_2020} studied a set of 15 stars, primarily solar twins, with ages in the range 27--4500\,Myr and activity indices $-4.8 < \log \RHK < -4.0$, in order to determine their large scale magnetic properties using optical intensity spectra. They found that magnetic regions have similar field strengths, $B = 3.2\pm0.6$\,kG, in all stars regardless of activity level; variations in the \textit{average} magnetic field $\langle B \rangle$ were instead driven by variations in the spot filling factor $\fspot$.
They found that power laws described well the relations between $\fspot$ and the mean magnetic field $\langle B \rangle$, and furthermore power law relations between $\langle B \rangle$ and several activity proxies including $\RHK$, across a wide range of levels of activity. 
We use their data to determine a power law fit, 
\begin{equation}
    \log \RHK = \log \fspot (0.77 \pm 0.19) - 3.82 \pm 0.14,
\end{equation}
and illustrate this in Fig.~\ref{fig:frac-lRHK}. 
The formal errors of the fit are shown by grey shaded regions in the plot, and range from 0.16\,dex at $\fspot = 0.4$ to 0.24\,dex at $\fspot = 0.1$. Conversely, at the solar activity level of $\log \RHK \approx -4.9$ \citep{noyes_rotation_1984} we find a $1 \sigma$ allowed range $0.01 < \fspot < 0.08$.
In our fit, we have excluded HD 29615, which lacks a reference $\log \RHK$ value, and $\xi$ Boo A, which is significantly cooler than the other stars in the sample and therefore may not follow the same spot--activity relationship.

Although not used in our tests, we also derive a relationship to the X-ray flux in the ROSAT \citep{boller_second_2016} 0.1--2.4\,keV band based on the data compiled by \citet{kochukhov_hidden_2020}, 
\begin{equation}
    \log L_X / L_\text{bol} = \log \fspot (2.43 \pm 0.53) - 2.71 \pm 0.39.
\end{equation}
This is advantageous for stars that may lack blue spectroscopy necessary to derive $\log \RHK$, 
but that may have X-ray measurements from large-sky surveys surveys such as (for the above data sources) ROSAT, Chandra \citep{evans_chandra_2010}, XMM-Newton \citep{webb_xmm-newton_2020}, or eROSITA \citep{merloni_srgerosita_2024}. Due to differences in the integrated energy band for integrated X-ray fluxes from different instruments, values should be converted to the ROSAT reference used here \citep[following, e.g.,][]{wright_stellar-activity-rotation_2011}.

We note that \citet{hahlin_determination_2023} performed a similar study to \citet{kochukhov_hidden_2020}, but using near-infrared rather than optical spectra. 
This enables measuring weaker magnetic fields thanks to the linear scaling of the Zeeman splitting with wavelength (eqn.~\ref{eqn:magsplit}).
Their results imply magnetic field strengths 2--3 times weaker than from optical data and spot filling factors 30\,\% smaller. 
Systematic errors beyond our statistical error bars may therefore be present in the translation from theoretical to observational parameter space.

\subsection{Predictions for spectroscopic effects in active stars}
\begin{figure*}
    \centering
    \includegraphics[width=\textwidth]{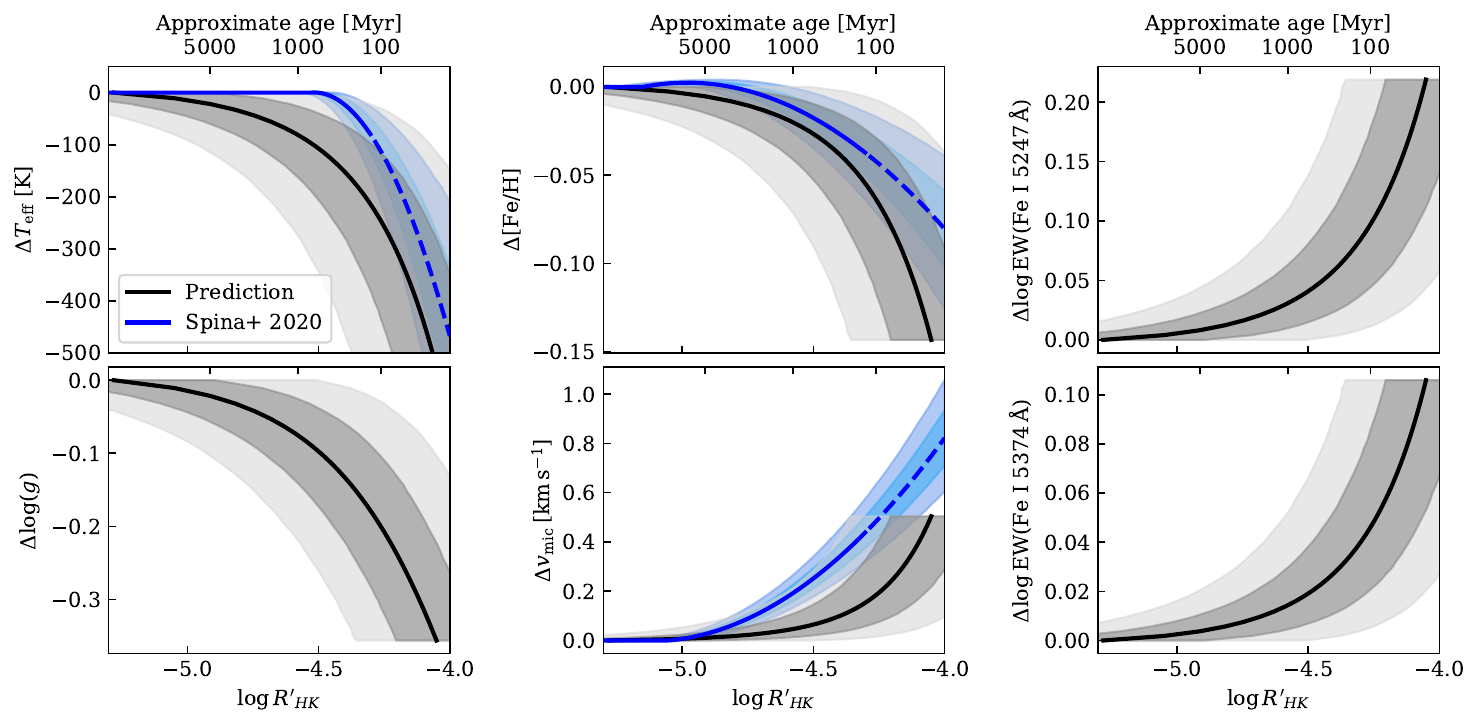}
    \caption{Predicted changes to stellar parameter measurements with variations in activity index, compared to empirically inferred effects for solar analogue stars \citep{spina_how_2020}. 
    The empirical effects have been extrapolated for $\RHK > -4.3$, and these regions are shown with dashed lines. 
    The $1\sigma$ and $2\sigma$ confidence regions are indicated by the dark and light shaded regions. For the predicted effects, uncertainties are estimated from the conversion from $\fspot$ into $\RHK$, and confidence regions are therefore horizontal. Approximate ages are indicated based on an empirical relationship between age and activity index \citep{mamajek_improved_2008}. }
    \label{fig:predictions}
\end{figure*}

We show in Fig.~\ref{fig:predictions} our predicted effects on stellar parameters for young solar analogues as a function of the activity index $\log \RHK$.
This uses the power law fit between $\log \RHK$ and $\fspot$, and with $\Tspot = 4771$\,K (i.e., adopting a typical umbral temperature 1000\,K below $\Teff$) and $\Bspot = 3$\,kG (a typical field strength for large magnetic regions). 
Predictions are compared to empirical estimates measured from solar twins  \citep{spina_how_2020}.
For clarity, an indicative age scale is also shown, based on a relationship between activity index and age \citep{mamajek_improved_2008}. 
We note that the more up to date $\log \RHK$--age relation from \citet{lorenzo-oliveira_solar_2018} by construction avoids the young and saturated regime, and thus is less applicable to the most active stars we aim to model here.

At an activity level of $\log \RHK = -4.3$, corresponding to an age of roughly 100\,Myr, we predict changes to the spectroscopic stellar parameters of 
$\Delta \Teff = -230$\,K,
$\Delta \logg = -0.20$\,dex, 
$\Delta \FeH = -0.06$\,dex, 
$\Delta \vmic = +0.17$\,\kms, 
and to the strengths of \ion{Fe}i \Felineone\ and \Felinetwo\ of
$+0.09$\,dex (15\,m\AA) 
and $+0.04$\,dex (7\,m\AA). 
We estimate error bars for these estimates based on uncertainties in the translation between $\log \RHK$ and $\fspot$, and these are roughly 140\,K, 0.10\,dex, 0.04\,dex, 0.14\,\kms, 0.06\,dex and 0.03\,dex. 
At the most extreme levels corresponding to $\log \RHK = -4.0$, these changes reach 
$\Delta \Teff = -700$\,K,
$\Delta \logg = -0.4$\,dex, 
$\Delta \FeH = -0.2$\,dex, 
$\Delta \vmic = +0.7$\,\kms, and for \ion{Fe}i \Felineone\ and \Felinetwo,
$+0.3$\,dex (60\,m\AA) 
and $+0.1$\,dex (25\,m\AA). 
Compared to \citet{spina_how_2020}, we predict an earlier but less steep onset for changes in $\Teff$, a later but more steep onset for $\vmic$, and very similar trends for $\FeH$. 
We note that \citet{spina_how_2020} didn't find significant variations in $\logg$, consistent with our modest predicted effect on this parameter. 

A potential conflict exists between our predicted effects on $\Teff$, which agree qualitatively with observations from \citet{spina_how_2020}, and the good agreement between spectroscopic and photometric temperatures found by \citet{baratella_gaia-eso_2020}. 
As indicated in Sect.~\ref{sec:results}, effects on the photometric $\Teff$ estimate are similar but not identical to the spectroscopic measurement. At $\log \RHK = -4.3$ and $-4.0$, we predict relatively small effects, $\Delta \Teff (B-V) = -100$ and $-300$\,K, and $\Delta \Teff (V-K_s) = -175$ and $-500$\,K. Disagreement at the 100\,K level should therefore be apparent in solar twins younger than 100\,Myr.
We note that while existing data clearly indicate brightness variations correlating with the stellar activity cycle \citep[e.g.,][]{hall_activity_2009}, variations in \textit{colour} are not well understood as existing data appear insufficient. 
Our predicted effects on colours are as large as 0.1\,mag for $B-V$ and 0.4\,mag for $V-K_s$, but we note that cyclical variations would be significantly smaller than this and therefore compatible with the lack of a signal in observations of $b-y$.

Our predicted effect on $\FeH$ at $\log \RHK = -4.3$ is in very close agreement with the average metallicity found in nearby star forming regions, $-0.07 \pm 0.03$\,dex \citep{cunha_chemical_1998,santos_chemical_2008,biazzo_chemical_2011,biazzo_chamaeleon_2012,dorazi_chemical_2011,spina_gaia-eso_2014,spina_gaia-eso_2017}. 
The agreement is very encouraging, but we note there may exist systematic differences between our spectroscopic analyses, not the least in the line selection.
Further, these stars are younger than 100\,Myr and may exhibit even stronger chromospheric activity. Our predicted effect on $\FeH$ at $\log \RHK = -4$ (age $\sim10$\,Myr) of $-0.15$\,dex would imply that these star forming regions actually have slightly supersolar metallicity \citep[in agreement with, e.g.,][]{baratella_gaia-eso_2020}.

\section{Conclusions and Summary} \label{sec:conclusions}
We have shown that star spots on stellar surfaces may systematically perturb spectroscopic measurements, in a way that depends critically on both the temperature and magnetic field of the spot. 
In tests on solar synthetic spectra, the temperature of the spot directly influences the inferred stellar effective temperature ($\Teff$), and cool spots may significantly strengthen certain spectral lines. 
For magnetically sensitive lines, Zeeman splitting effectively produces broadening with a width similar to the intrinsic line width. For saturated lines, this leads to an amplification that partly mimics the effects of microturbulence ($\vmic$), but varies from line to line according to their magnetic sensitivity ($\geff$). 

We have analysed synthetic spectra of spotted stars with solar parameters in a detailed line-by-line differential spectroscopic analysis, treating them as if they were real stars. 
When modelling strongly magnetic cool spots, we infer systematic changes in the spectroscopic stellar parameters that vary directly with the surface spot coverage ($\fspot$). We find an interplay between the spot temperature and magnetic field strength that strengthens the effects on the inferred $\Teff$, $\logg$ and $\FeH$, but partly cancels for effects on $\vmic$. 
When adopting a relation between the $\log \RHK$ activity index and $\fspot$, we find systematic variations in stellar parameters that are in agreement with observations of young stellar twins \citep{spina_how_2020}. We find remarkably good agreement in the sign and magnitude of effects, but mild disagreement in the activity level where these effects set in. 
Our simulations indicate an onset of significant effects on spectroscopic stellar parameters for stars with $\log \RHK \gtrsim -4.6$ corresponding to ages less than about 1\,Gyr \citep{mamajek_improved_2008}. Effects grow stronger for more active, younger, stars. 

Although we predict effects in good agreement with observational data from \citet{spina_how_2020}, we note that the effects on $\Teff$ are predicted to onset at too low activity levels, while effects on $\vmic$ appear only at high activity levels. It is unlikely these differences could be explained solely by systematic uncertainties or variations in the empirical $\log \RHK$--$\fspot$ relationship. Rather, they may reflect shortcomings in the observational data, our simple model of stars with cool spots, or both.
We note also that our inferred effects may not be directly applicable to studies that used other linelists, nor to stars that are not solar analogues.

The empirical $\log \RHK$--$\fspot$ relationship used in this work was derived from measurements of magnetic spots in young solar analogues \citep{kochukhov_hidden_2020}. It should be verified whether this relationship applies also to stars of other spectral types, and further how parameters like the magnetic field strength or the spot temperature (contrast) depends on the spectral type.

Additional data on young stars throughout their activity cycle, and especially for stars with cycles that span a wide range in the activity index at a high average value, will help better constrain the empirical effects. 
Data from \citet{spina_how_2020} covers solar twins over a wide range of activity indices, but is least complete for moderately-to-very active stars with $\log \RHK$ values in the range $-4.7$ to $-4.0$ (ages $<2$\,Gyr); this is also the range where effects are significant. 
Their data do not indicate star-to-star variations, but rather agree well with a simple dependence on the activity level.

A potentially important modelling shortcoming is that the presence of strong magnetic significantly alters the structure of the photosphere beyond the assumptions of our simple model. 
Advanced 3D MRHD simulations have been used to predict photospheric structures in a small number of cases for the Sun \citep[e.g.][]{heinemann_mhd_2007,rempel_penumbral_2009,rempel_radiative_2009,rempel_penumbral_2011,rempel_numerical_2012} and stars of different spectral types \citep[][]{panja_3d_2020}. 
Due to the immense computational cost of such simulations, it is likely that a simplified approach to describe magnetic spots is required for detailed applications such as abundance analyses.
Another mechanism, further to that considered in this work, is that chromospheric emission may be present in the cores of strong lines that form in the highest layers of the photosphere. Semi-empirical chromospheric model grids like that presented by \citet{vieytes_chromospheric_2005} could be used to test this effect.
We note also that the use of hydrostatic non-magnetic model atmospheres with adjusted effective temperature, although common in the study of spotted stars, may be insufficient for detailed abundance work and indeed is not the standard in solar physics. Semi-empirical model atmospheres such as the umbral model of \citet{maltby_new_1986} typically exhibit a flatter temperature structure and an overall lower gas pressure than standard hydrostatic atmospheres, which would alter the spectral imprint of spots.

Large cool and warm spot structures have been seen on young solar analogues, e.g., V1358 Ori \citep{kriskovics_magnetic_2019}. It's not clear if the warm spots in these cases are analogous to cool spots, or if they are large versions of faculae which are seen on the solar surface, and which are known to dominate over cool spots in stars of solar age \citep[e.g.,][]{radick_patterns_1998,montet_long-term_2017}. 
Regardless, such findings may call for a more involved model with additional parameters, e.g., $\Delta T_{\rm cool}$, $\Delta T_{\rm hot}$, $B_{\rm cool}$, $B_{\rm warm}$, $f_{\rm warm}$ and $f_{\rm warm}$. 
Empirical data on these parameters are necessary in order to constrain such a model, especially if the warm and cool spot coverage fractions vary differently with age. We therefore caution the reader  that in case the presence of warm spots is indeed crucial, then the model proposed in this work will be insufficient in this regard.
In contrast, our current model is sufficiently simple and straightforward that it can be implemented in stellar spectroscopy packages with the addition of the single additional stellar parameter $\log \RHK$ (or $\log L_X / L_\text{bol}$), and can be used without any changes to the spectroscopic setup.

Further work on these effects will focus on improving the theoretical model and extending the application to young dwarfs and giants more broadly. A full treatment of polarised radiative transfer is likely necessary in order to accurately represent the effects of Zeeman splitting \citep[see, e.g.,][]{cristofari_constraining_2023}. 
We will include these effects in our future work, where we also focus on outstanding problems like the Ba puzzle \citep{dorazi_enhanced_2009}.

\section*{Acknowledgements}
We thank the anonymous referee for their thorough reading of the manuscript, which led to a large number of improvements.
Parts of this research were supported by the Australian Research Council Centre of Excellence for All Sky Astrophysics in 3 Dimensions (ASTRO 3D), through project number CE170100013.
This work was supported by computational resources provided by the Australian Government through the National Computational Infrastructure (NCI) under the National Computational Merit Allocation Scheme (project y89). V.D. acknowledges the financial contribution from PRIN MUR 2022 (code 2022YP5ACE) funded by the European Union – NextGenerationEU.

\section*{Data Availability}
The radiative transfer code TurboSpectrum is freely available (\citealt{plez_turbospectrum:_2012}; see also \citealt{gerber_non-lte_2023}), and the molecular linelists (Table~\ref{tbl:molecules}) are available for download from Bertrand Plez's website \url{https://www.lupm.in2p3.fr/users/plez}. 
The linelist modifications for Zeeman splitting implemented here are trivial, but the code will be shared upon request. 
The spectrum analysis codes are likewise freely available: 
\texttt{smhr}, \texttt{stellar diff} and \texttt{qoyllur-quipu} (q$^2$)
are available at 
\url{https://github.com/andycasey/smhr},
\url{https://github.com/andycasey/stellardiff}
and \url{https://github.com/astroChasqui/q2}.




\bibliographystyle{mnras}
\bibliography{zotero-library,extra_refs}

\appendix
\section{Spectrum synthesis linelist bibliography}
The sources of molecular linelists are provided in Table~\ref{tbl:molecules}.

\begin{table}
    \caption{Complete list of molecular line lists used in synthetic spectrum calculations in this work.}
    \begin{tabular}{l c}
         Species & References \\ \hline 
         AlH & Plez \& Jorissen (priv. comm.)\\
         $\rm C_2$ &  \citet{brooke_line_2013,ram_improved_2014} \\
         CaH & Plez (priv. comm.) \\
         CH &  \citet{masseron_ch_2014} \\
         CN &  \citet{brooke_einstein_2014,sneden_line_2014} \\
         CO &  \citet{goorvitch_infrared_1994} \\
         CrH & \citet{burrows_new_2002} \\ 
         FeH & \citet{dulick_line_2003} \\
         $\rm H_2O$ & \citet{barber_high-accuracy_2006} \\
         MgH & \citet{skory_new_2003} \\ 
         NH  & \citet{kurucz_kurucz_1995-1}\\ 
         OH &  \citet{kurucz_kurucz_1995-1}, \citet{goldman_updated_1998}\\ 
         SiH & \citet{kurucz_kurucz_1995-1}\\ 
         SiO & \citet{langhoff_theoretical_1993} \\ 
         SiS & \citet{cami_detection_2009} \\ 
         TiO & \citet[][updated Jan 2012]{plez_new_1998} \\
         VO & Plez (priv. comm.) \\ 
         ZrO & Plez (priv. comm.) \\
    \end{tabular}
    \label{tbl:molecules}
\end{table}

\bsp	
\label{lastpage}
\end{document}